%% file: manuscript.tex
\documentclass{jfm}
\input{pream}

\DeclareRobustCommand\full  {\tikz[baseline=-0.6ex]\draw[thick] (0,0)--(0.5,0);}
\DeclareRobustCommand\dotted{\tikz[baseline=-0.6ex]\draw[thick,dotted] (0,0)--(0.54,0);}
\DeclareRobustCommand\dashed{\tikz[baseline=-0.6ex]\draw[thick,dashed] (0,0)--(0.54,0);}
\DeclareRobustCommand\dashdot {\tikz[baseline=-0.6ex]\draw[thick,dash dot] (0,0)--(0.5,0);}

\definecolor{midGreen}{rgb}{0., 0.5, 0.}
\definecolor{redBlue}{rgb}{0.5, 0., 0.5}

\graphicspath{{./figures/}{./figures/phdcomics/}}

\newcommand{\tmpFigHeight}{0.17\textheight}
\newcommand{\tmpFigWidth}{0.33\textwidth}
\shortauthor{S. Vadarevu, S. Illingworth, and I. Marusic}

\title{Coherent structures in the linearized impulse response of turbulent channel flow}

\author{Sabarish B. Vadarevu
\corresp{\email{sabarish.vadarevu@unimelb.edu.au}},
  Simon J. Illingworth
 \and Ivan Marusic }

\affiliation{Department of Mechanical Engineering, The University of Melbourne,
Parkville, VIC 3010, Australia
}
\begin{document}
\maketitle

\begin{abstract}
    We study the evolution of velocity fluctuations due to an isolated spatio-temporal impulse using the linearized Navier-Stokes equations. The impulse is introduced as an external body force in incompressible channel flow at $Re_\tau=10000$. Velocity fluctuations are defined about the turbulent mean velocity profile. A turbulent eddy viscosity is added to the equations to fix the mean velocity as an exact solution, which also serves to model the dissipative effects of the background turbulence on large-scale fluctuations. 
    An impulsive body force produces flowfields that evolve into coherent structures containing long streamwise velocity streaks that are flanked by quasi-streamwise vortices; some of these impulses produce hairpin vortices. As these vortex-streak structures evolve, they grow in size to be nominally self-similar geometrically with an aspect ratio (streamwise to wall-normal) of approximately $10$, while their kinetic energy density decays monotonically. The topology of the vortex-streak structures is not sensitive to the location of the impulse, but is dependent on the direction of the impulsive body force. All of these vortex-streak structures are attached to the wall, and their Reynolds stresses collapse when scaled by distance from the wall, consistent with Townsend's attached eddy hypothesis. 
\end{abstract}

\input{introduction}
\input{formulation}
\input{impres}

\input{outlook}

\bibliographystyle{jfm}
\bibliography{biblio}
\end{document}

%% file: pream.tex
\usepackage{textcomp}
\usepackage{graphicx}		
\usepackage[utf8]{inputenc}
\usepackage[T1]{fontenc}
\usepackage{hyperref}

\usepackage{amsmath}		
\usepackage{amsfonts}
\usepackage{amssymb}
\usepackage{bm}
\usepackage{morefloats}
\usepackage{mathtools}
\usepackage{setspace}
\usepackage{appendix}
\usepackage{cleveref}

\usepackage{tikz}
\usetikzlibrary{shapes.geometric, arrows}
\crefname{section}{\S}{\S\S}
\Crefname{section}{\S}{\S\S}

\usepackage{pgfplots}
\pgfplotsset{compat=newest,ticklabel style = {font=\tiny}}
\usetikzlibrary{backgrounds,automata}

\numberwithin{equation}{section}

%% file: introduction.tex
\section{Introduction}
    Linear processes play an essential role in the subcritical transition to turbulence \citep{henningson1994role,waleffe1995transition} and to the sustenance of turbulence \citep{kim2000linear} in wall-bounded shear flows. Linear analyses involve decomposing the velocity fields into a base flow and perturbations about the base flow; for transition to turbulence, the base flow is the steady laminar solution, while for fully turbulent flows, the appropriate choice for the base flow is the turbulent mean velocity \citep{reynolds1967stability}. The turbulent mean velocity is linearly stable, and the coupling term in the linearized Navier-Stokes equations  (LNSE) is responsible for the sustenance of turbulence structures as shown by \cite{kim2000linear}. In more recent work, \cite{mckeon2010critical} exposed the nature of the linear processes to be that of selective amplification of certain modes (which are travelling waves in their analysis), which allows a substantial reduction in the dimensionality (number of degrees of freedom) of high Reynolds number turbulence \citep{moarref2013model}.     
    
    A turbulent eddy viscosity is often introduced to the LNSE to fix the turbulent mean velocity profile as an exact solution of the equations, and to model the dissipative effect of the small-scales of turbulence on the large-scale perturbations \citep{reynolds1972mechanics}. The eddy-viscosity-enhanced linearized Navier-Stokes equations (eLNSE) have been successful in describing several features of wall-bounded turbulence. For instance, \cite{del2006linear} and \cite{pujals2009note} showed that dominant turbulence structures are captured by the eLNSE as the perturbations with the largest transient growth; \cite{hwang2010linear} extended these to study the response to harmonic and stochastic forcing, and showed the existence of inner and outer peaks in spanwise size of maximally amplified structures, as well as a self-similar range between the two peaks. \cite{hwang2016mesolayer} suggested that the eddy viscosity enhancement could also help reconcile the inner-outer interaction \citep{mathis2009large} in the near-wall region. A review of LNSE and eLNSE models for wall-bounded turbulence has recently been presented by \cite{mckeon2017engine}.

The impulse response completely characterizes a linear dynamical system in the time domain. In the context of the linearized dynamics of turbulence, this involves studying the response to non-linearities that are localized in both space and time. Using the impulse response to study coherent structures has the advantage of parametrizing their evolution without the need to impose \textit{a priori} either their wall-parallel dimensions (i.e. fundamental Fourier modes in the streamwise and spanwise directions) or a phase relationship between the Fourier modes that constitute these structures. Several researchers have applied the impulse response or similar tools to turbulent flows. Although not using the impulse response directly, \cite{landahl1990sublayer} argued that the near-wall dynamics of turbulence can be considered as the linear response to spatio-temporally localized non-linear terms because of the high intermittency of wall-normal velocity fluctuations. \cite{jovanovic2001spatio} used the impulse response to study the evolution of perturbations about a laminar base flow. \cite{luchini2006phase} did this for turbulent flows by introducing white noise at the wall in a DNS and computing space-time correlations at $Re_\tau = 180$. \cite{codrignani2014impulse} extended the DNS-based study to different wall-normal locations for the impulse at a slightly lower Reynolds number of $Re_\tau=150$. More recently, \cite{eitel2015hairpin} looked at the evolution of hairpin vortices at a higher Reynolds number of $Re_\tau =590$, using the eLNSE as well as DNS to include a fully turbulent background. They found that the evolution of a single hairpin vortex observed in the turbulent DNS is well represented by the eddy-viscosity-enhanced LNSE; however, the linear model failed to capture regeneration, consistent with the earlier observations of \cite{kim2000linear}.  
    
    In this paper, the low computational cost of the eLNSE model is exploited to extend these investigations of the impulse response to turbulent flows at higher Reynolds numbers; we use $Re_\tau = 10000$ to allow for a decade of wall-normal extent of the log layer, although computations at higher $Re_\tau$ are also feasible. The aims of the work are to investigate (i) if the impulse response captures the vortex-streak structure known to be important to turbulence \citep{jimenez1991minimal,hwang2015statistical}, (ii) the prevalence of the hairpin vortex \citep{adrian2007hairpin}, and (iii) the geometrical scaling of the coherent structures and associated turbulent stresses as they evolve in time. The numerical formulation for computing the impulse response is presented in \cref{sec:formulation}, and the results for nine different cases of impulses are shown in \cref{sec:results}. In \cref{sec:outlook}, we summarize the principal observations of this paper and discuss the limitations of the present model. 

%% file: formulation.tex
\section{Numerical formulation}\label{sec:formulation}
The flow under consideration is incompressible channel flow. 
The velocity field is decomposed as the turbulent mean and perturbations about this mean, $\mathbf{u}_{total} = \lbrack U(z), 0,0 \rbrack + \mathbf{u}$. A Cartesian coordinate system is employed, where $x$, $y$, and $z$ represent the streamwise, spanwise, and wall-normal directions respectively. The perturbation vector $\mathbf{u}$ has components $\lbrack u,v,w\rbrack^T$, where $u$ is the streamwise velocity, $v$ is the spanwise velocity, and $w$ is the wall-normal velocity. We are interested in the evolution of these perturbations about a specified turbulent mean. The turbulent mean, $U(z)$, is computed by integrating an integrand containing the turbulent eddy viscosity,  
$(-u_\tau Re_\tau \eta)/  (\nu_T/\nu) $,
from $\eta=-1$ to $z$. Here, $z$ is the wall-normal coordinate non-dimensionalized by channel half-height such that the channel walls are at $z = \pm 1$, $\eta$ is a proxy for the non-dimensionalized wall-normal coordinate, $Re_\tau$ is the Reynolds number based on the friction velocity $u_\tau$, and $\nu_T$ is the total eddy viscosity. The total eddy viscosity $\nu_T=\nu_t+\nu$ contains the turbulent eddy viscosity $\nu_t$ and molecular viscosity $\nu$, and is approximated using the semi-analytical expression of \cite{cess1958survey} as reported by \cite{reynolds1967stability}: 
\begin{equation}\label{eddy-viscosity}
    \nu_T = \frac{\nu}{2}\bigg\lbrack 1 + 
            \frac{\kappa^2 Re^2_\tau}{9} (1-z^2)^2 ( 1+2z^2 )^2 
            \bigg\{ 1 - \exp\bigg( \frac{Re_\tau (|z| -1 )}{A}\bigg) \bigg\}^2
            \bigg\rbrack ^{1/2} + \frac{\nu}{2},
\end{equation}
where values for the von K\'arm\'an constant $\kappa =0.426$, and the constant $A=25.4$ are set to fit the turbulent mean velocity profile at $Re_\tau = 2000$ \citep{hoyas2006scaling}.  

We investigate the evolution of perturbations due to a spatio-temporally impulsive body force, which is considered to represent an idealized bursting event. This spatio-temporal impulse, $\tilde{\delta}(\mathbf{x}-\mathbf{x}_0,t)$, is factorized into impulses along each direction and time as
\begin{equation}
     \tilde{\delta}(\mathbf{x}-\mathbf{x}_0,t) = \delta(x)\delta(y) \delta(z-z_0)  \delta(t),
\end{equation}
where $\delta$ is the Dirac delta function. The impulse is introduced at $x=y=t=0$, with a variable wall-normal location $z_0$. The Fourier transform of the impulse along $x$ and $y$ produces a coefficient of unity for each Fourier mode. 

A momentum impulse along the direction $( m_x \hat{\mathbf{e}}_x+ m_y \hat{\mathbf{e}}_y +  m_z\hat{\mathbf{e}}_z )$, $\hat{\mathbf{e}}_s$ being the unit-vector along direction $s\in \{x,y,z\}$, enters the evolution equations for any Fourier mode $e^{i(k_x x + k_y y)}$ as follows. 
\begin{equation}\label{impulse-system}
    \begin{aligned}
        \partial_t \hat{\boldsymbol{\phi}} &= 
            A\hat{\boldsymbol{\phi}}  + 
                B\hat{\mathbf{f}} \delta(t)  ,\\
        \hat{\mathbf{u}} &= C\hat{\boldsymbol{\phi}} ,
    \end{aligned}
\end{equation}
where the state $\hat{\boldsymbol{\phi}}=\lbrack \hat{w} , \hat{\eta} \rbrack^T$ contains the Fourier coefficients of wall-normal velocity and wall-normal vorticity, and $\hat{\mathbf{f}}(z) = \lbrack m_x \delta(z-z_0) ,m_y \delta(z-z_0) ,m_z \delta(z-z_0)\rbrack^T $ is the wavenumber-independent Fourier coefficient for the impulsive body force (which models an idealized bursting event in the present work). The operators $A$, $B$, and $C$ are dependent on the wavenumbers of the Fourier mode, and are presented below. 
\begin{equation}
    \begin{gathered}
    A =
        \begin{bmatrix} \Delta^{-1} \mathcal{L}_{OS}  & 0 \\ 
        -ik_y U' & \mathcal{L}_{SQ}      \end{bmatrix},\quad
    C = \frac{1}{k_x^2 + k_y^2} \begin{bmatrix}
        ik_x \mathcal{D} & -ik_y \\
        ik_y\mathcal{D}   & ik_x \\
        k_x^2 + k_y^2  & 0 
        \end{bmatrix},
        \\
    B = 
        \begin{bmatrix}
            -ik_x \Delta^{-1} \mathcal{D} & -ik_y\Delta^{-1}\mathcal{D} & -(k_x^2 + k_y^2)\Delta^{-1}  \\
        ik_y  & -ik_x & 0     \end{bmatrix},
    \end{gathered}
\end{equation}
where, the prime and $\mathcal{D}$ represent wall-normal differentiation, $\Delta = \mathcal{D}^2 - (k_x^2 +k_y^2)$, and the operators $\mathcal{L}_{OS}$ and $\mathcal{L}_{SQ}$ are
\begin{equation}
    \begin{aligned}
        \mathcal{L}_{OS} &= ik_x U'' - ik_x U \Delta + \nu_T \Delta^2 + 2 \nu'_T \Delta \mathcal{D} + \nu_T''\big(\mathcal{D}^2 + (k_x^2 + k_y^2)\big),\\
        \mathcal{L}_{SQ} &= -ik_x U + \nu_T \Delta + \nu'_T \mathcal{D} .
    \end{aligned}
\end{equation}
See chapter 10 of \cite{jovanovic2004modeling} for a similar investigation of perturbations about the laminar flow; the operators $B$ and $C$ remain the same as in that study, while the operator $A$ is identical to the one used in \cite{hwang2010linear} to reflect the linearization about the turbulent mean velocity and the introduction of the turbulent eddy viscosity.

The velocity perturbations for any Fourier mode at a time $t$ due to an impulse introduced at time $t=0$ is calculated explicitly for any time $t>0$, without resorting to time-marching, as 
\begin{equation}\label{impulse-response}
    \hat{\mathbf{u}} = C e^{At} B \hat{\mathbf{f}} ,
\end{equation}
which is obtained by integrating \cref{impulse-system}. The impulse response is computed for a spatially periodic flow with domain sizes $8\pi$ and $3\pi$ along the streamwise and spanwise directions. The Fourier modes used to construct the field of perturbations are integral multiples of the fundamental wavenumbers $k_{x,0}=(2\pi)/(8\pi)$ for streamwise and $k_{y,0}=(2\pi)/(3\pi)$ for spanwise directions. The modes are truncated so that higher wavenumbers that are excluded have less than $0.001\%$ of the energy in the most energetic Fourier mode; the number of Fourier modes needed to meet this truncation criterion goes from $(160,192)$ modes along streamwise and spanwise directions (counting positive and negative wavenumbers) at early times to $(40,40)$ modes at later times. Along the wall-normal direction, the Chebyshev collocation method is used.

To facilitate consistent discretization, the wall-normal Dirac delta function is approximated by the exponential function,\\
\begin{equation}\label{fs0}
    \delta(z-z_0) \approx f_{s0}(z;z_0) := \frac{K}{2\sqrt{\pi \epsilon}} \exp\bigg(-\frac{\big(Re_\tau(z-z_0)\big)^2}{4\epsilon}\bigg),
\end{equation}
where $z_0$ is the location of the impulse, and $\epsilon$ quantifies the width of the impulse. We use $\epsilon = z^+_0/4$ to have consistent discretization of the exponential function in \cref{fs0} near the wall and in the core of the channel; here, $z^+ = Re_\tau(1+z)$ is the wall-normal coordinate expressed in inner units and referenced from the bottom wall, so that the bottom wall is at $z=-1$ and $z^+=0$. The constant $K$ in \cref{fs0} is set so that the function has an area of unity. The number of Chebyshev nodes used in the wall-normal direction is $768$; increasing this number to $1152$ produces a relative change in energy of less than $0.01\%$ for all Fourier modes and times considered here.

The operators $A,B$, and $C$ have been validated against the results of \cite{pujals2009note} and \cite{hwang2010linear}. The impulse response was separately validated about a laminar base flow against the results in chapter 10 of the PhD thesis of \cite{jovanovic2004modeling}. 

%% file: impres.tex
\section{Results}\label{sec:results}
\renewcommand{\tmpFigHeight}{0.09\textheight}
\renewcommand{\tmpFigWidth}{\textwidth}
\begin{figure}
    \centering
    \includegraphics[height=\tmpFigHeight,width=\tmpFigWidth,trim={3cm 0 2cm 0},clip]{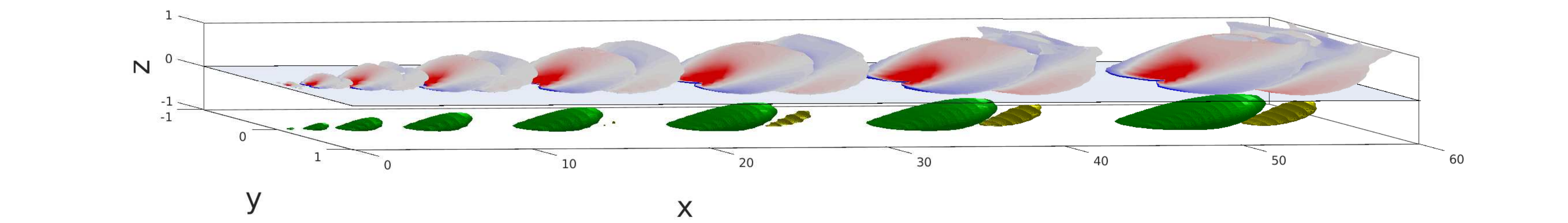}\\
    \includegraphics[height=\tmpFigHeight,width=\tmpFigWidth,trim={3cm 0 2cm 0},clip]{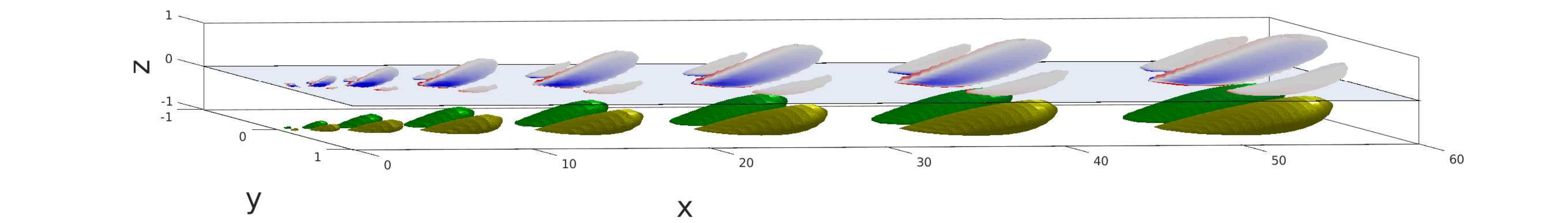}\\
    \includegraphics[height=\tmpFigHeight,width=\tmpFigWidth,trim={3cm 0 2cm 0},clip]{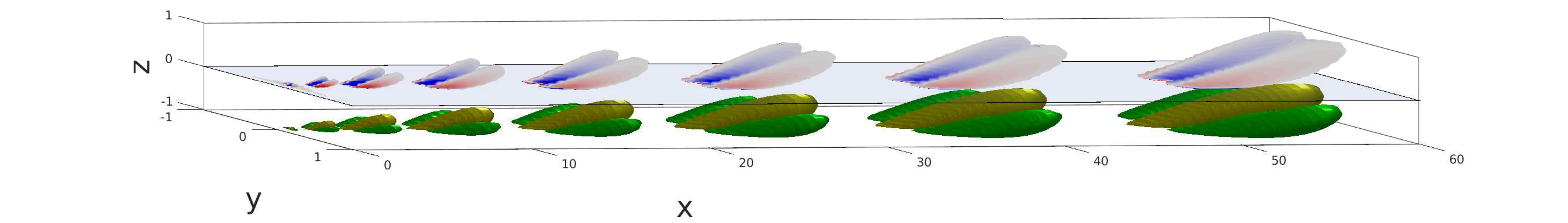}
    \caption{Coherent structures at times (from left to right) $tU_{CL}/h$ from $1.42$ to $41.2$ in steps of $5.7$ due to the impulses (from top to bottom) I30x, I30y, and I30z. Green-yellow isosurfaces are for streamwise perturbation velocity at $+25\%$ of instantaneous maximum (green) and $-25\%$ of instantaneous maximum (yellow), with their streamwise locations modified to facilitate visualization. Red-white-blue isosurfaces are for swirling strength at $10\%$ of the instantaneous maximum (colored by spanwise vorticity), and are plotted with a wall-normal offset of 1 to avoid overlap with the streamwise velocity isosurfaces.\label{fig:swirl-lapse-full}}
\end{figure}
\renewcommand{\tmpFigWidth}{0.33\textwidth}
\begin{figure}
    \centering
    \includegraphics[width=\tmpFigWidth,height=\tmpFigHeight]{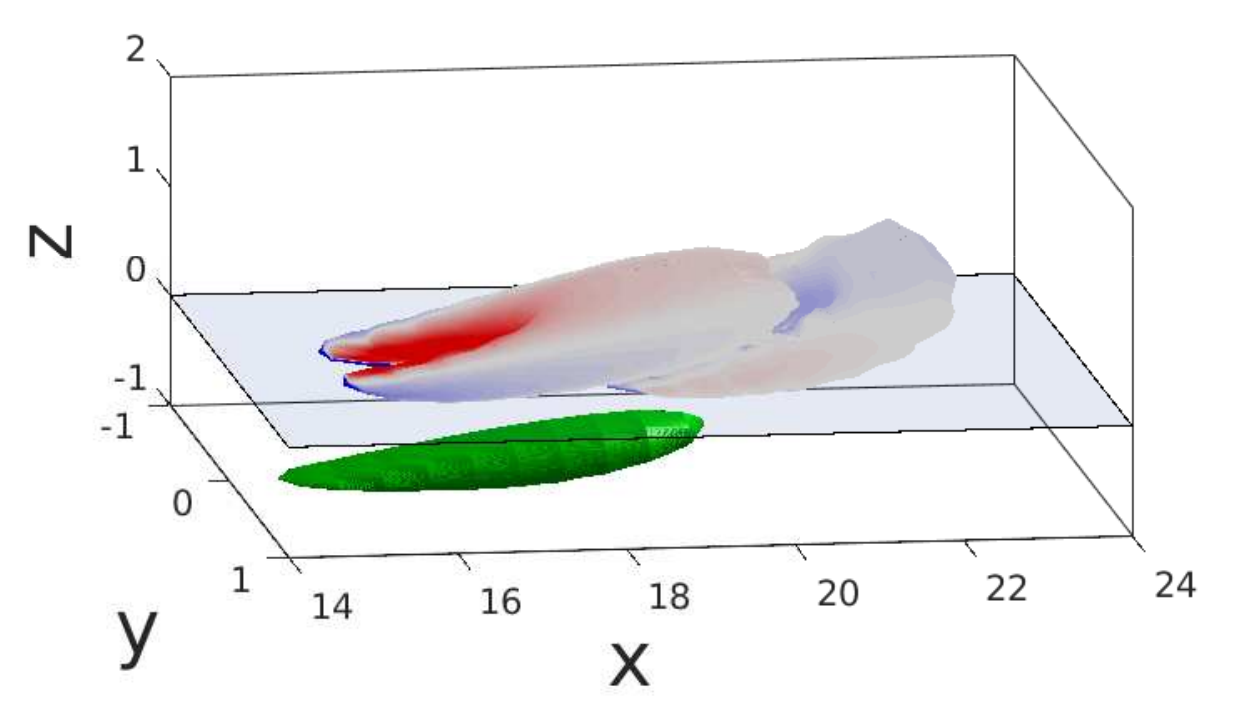}\hfill
    \includegraphics[width=\tmpFigWidth,height=\tmpFigHeight]{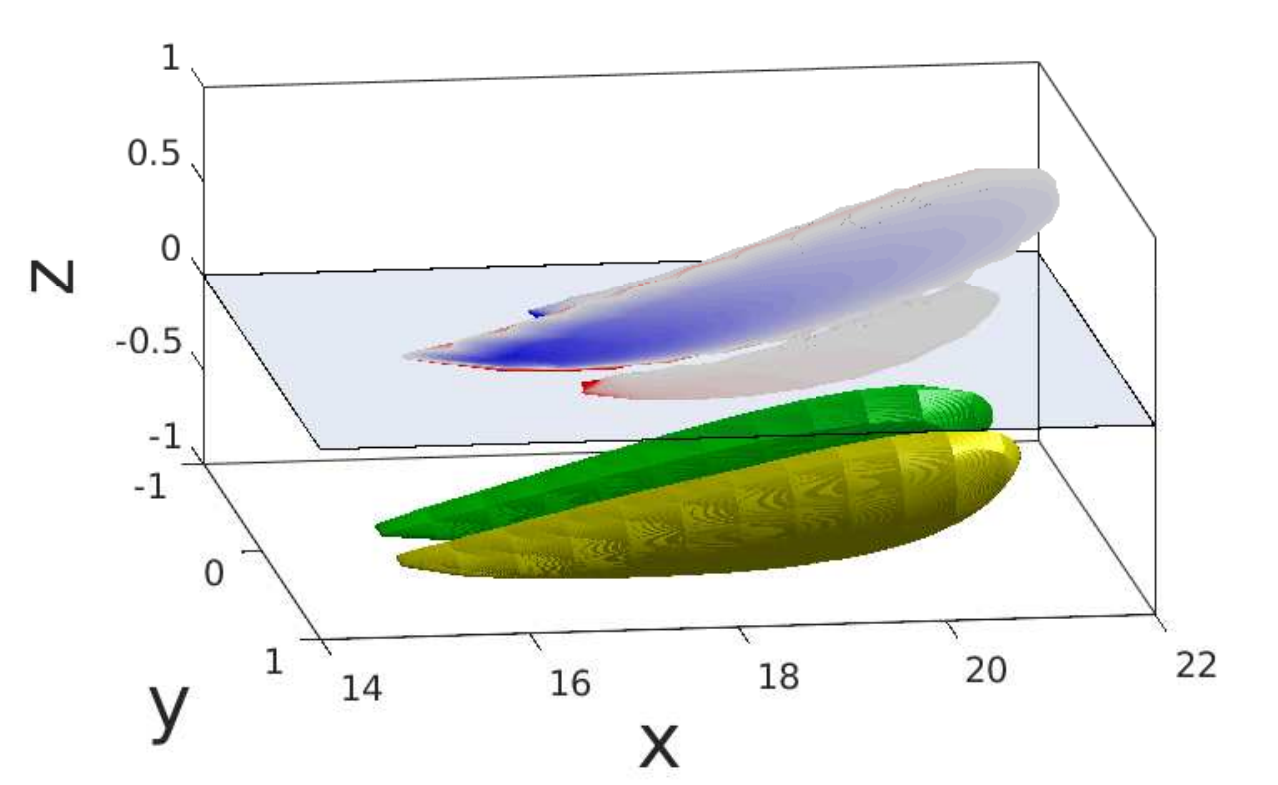}\hfill
    \includegraphics[width=\tmpFigWidth,height=\tmpFigHeight]{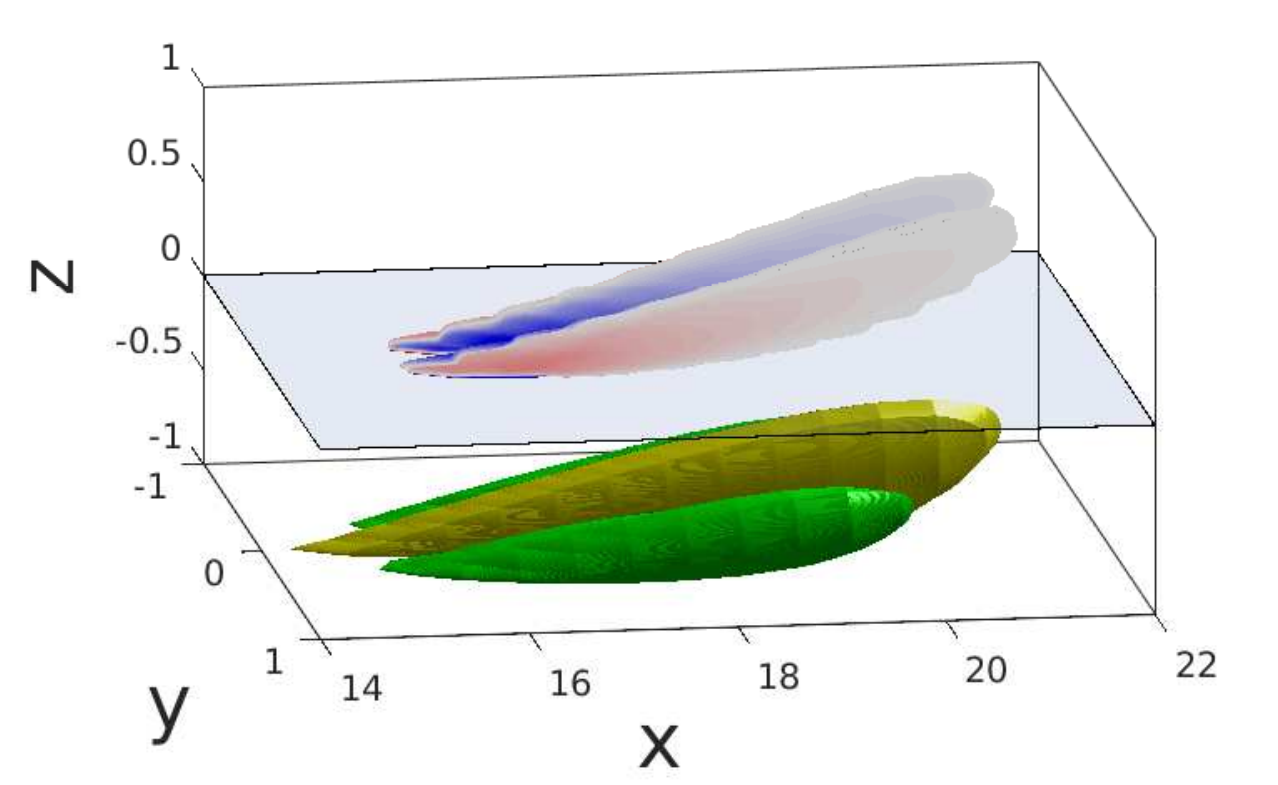}\\
    \includegraphics[width=\tmpFigWidth,height=\tmpFigHeight]{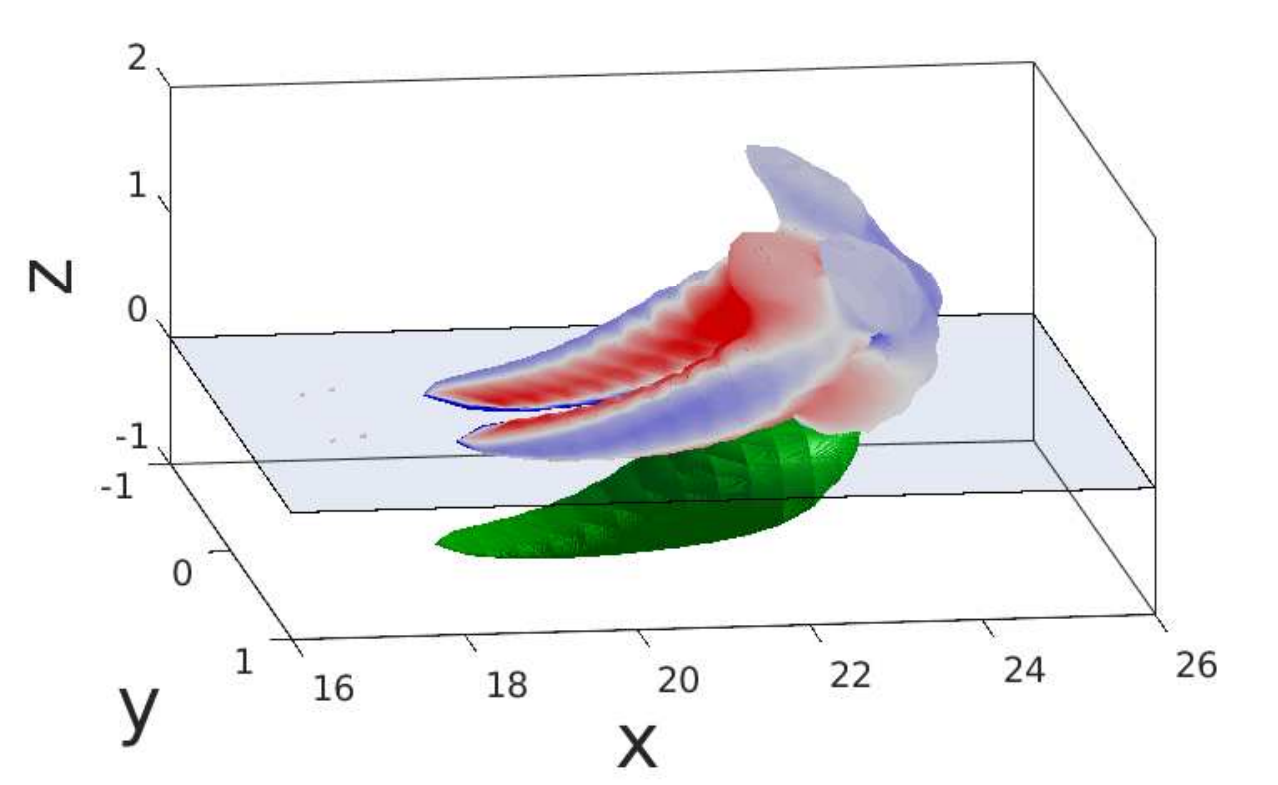}\hfill
    \includegraphics[width=\tmpFigWidth,height=\tmpFigHeight]{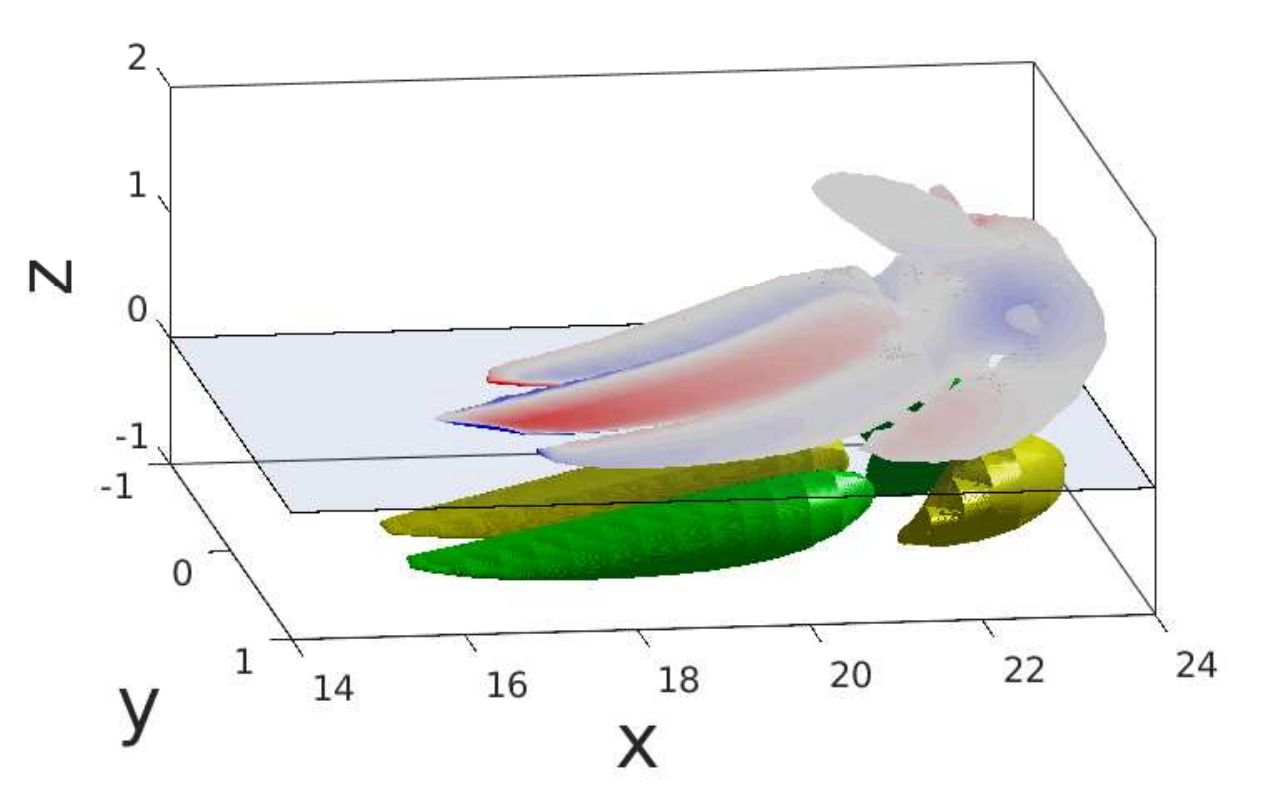}\hfill
    \includegraphics[width=\tmpFigWidth,height=\tmpFigHeight]{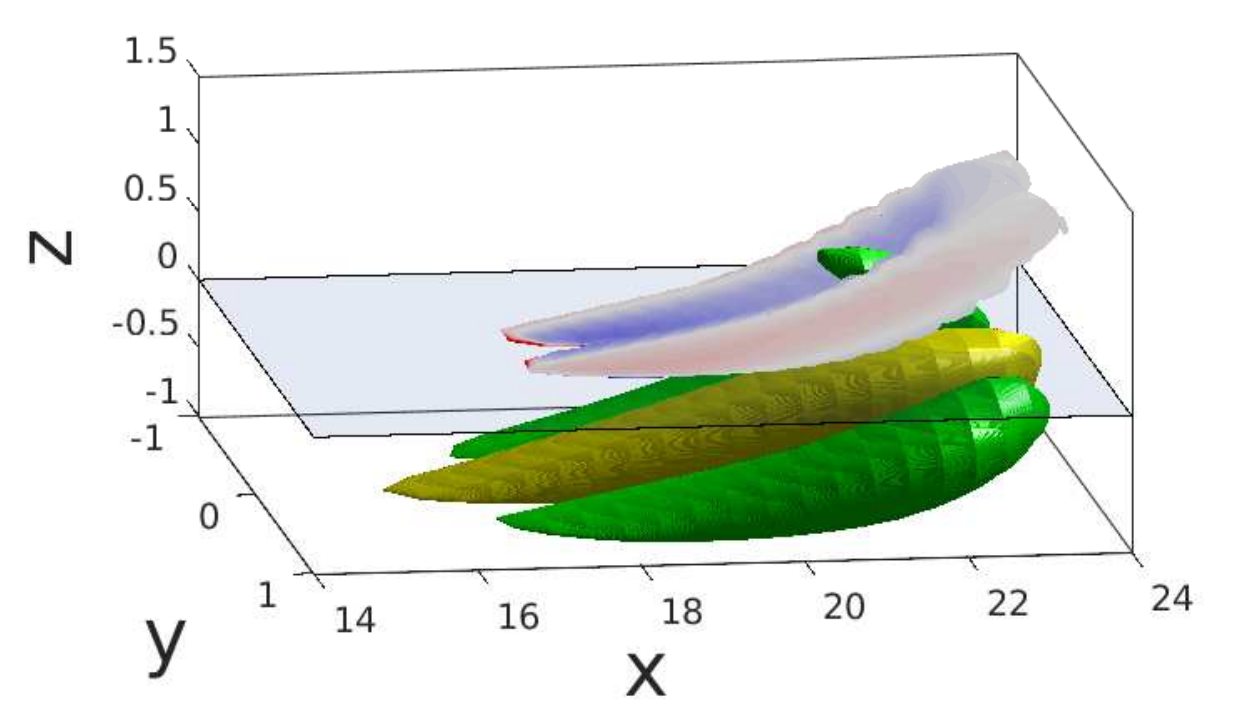}
    \caption{Coherent structures due to the impulses at $z^+=500$ (top) and $z=-0.5$ (bottom) along directions $x$ (left), $y$ (middle), and $z$ (right) at time $tU_{CL}/h = 24.1$. Isosurfaces are as in \cref{fig:swirl-lapse-full}, and swirling strength has a wall-normal offset of 1 as before to avoid overlap with velocity isosurfaces.\label{fig:swirl-inst-full}}
\end{figure}

The response to impulses introduced at three different locations, $z^+ \approx 30$, $z^+\approx 500$, and $z = -0.5$ (or $z^+=5000$) at $Re_\tau = 10000$ were computed at times $tu_{\tau}/h$ from $0.05$ to $1.5$ in steps of $0.05$, or $tU_{CL}/h$ from $1.42$ to $41.2$ in steps of $1.42$. At each location, the body force is introduced along each of the three coordinate axes, producing a total of nine cases. For convenience, these cases shall be identified with a name such as I30x, where the `I30' part refers to the forcing location being 30 inner units, and `x' refers to the forcing direction. The names for the first six cases shall start with `I30' or `I500', followed by the direction of the forcing, while the last three cases shall start with `O05' followed by the direction. 

The topology of the coherent structures produced by the impulse response at different times is illustrated in \cref{fig:swirl-lapse-full}, where isosurfaces of swirling strength and streamwise velocity are plotted for I30x, I30y, and I30z cases for 8 different times: $tU_{CL}/h$ from $1.42$ to $41.2$ in steps of $5.7$. The same isosurfaces are shown in \cref{fig:swirl-inst-full} for the I500 and O05 cases, but at only one time instant, $tU_{CL}/h=24.1$. The swirling strength is defined at each point in the domain as the imaginary part of the complex eigenvalue of the velocity gradient tensor \citep{zhou1999mechanisms}. The coherent structure produced by the impulse contains streamwise velocity streaks which are flanked or enclosed by quasi-streamwise vortices. The coordinate axes are scaled unequally to facilitate visualization; the streamwise size of the structures is about 10 times longer than their wall-normal or spanwise size. The topology of the streaks and vortices change when the direction of the body force is changed, but changing the location from $z^+\approx 30$ to $z^+ \approx 500$ does not affect a significant change. Forcing at $z = -0.5$ produces a more observable change in the structure of the perturbations, especially when the structures extend into the top half of the channel; however, features such as the number of streaks and vortices, and their relative arrangement, are retained. All of the structures also seem to be attached to the wall; this becomes more clear when Reynolds stresses are plotted. Geometric self-similarity is nominally observed in the evolution of these vortex-structures.

\begin{figure}
    \centering
    \includegraphics[width=0.99\textwidth]{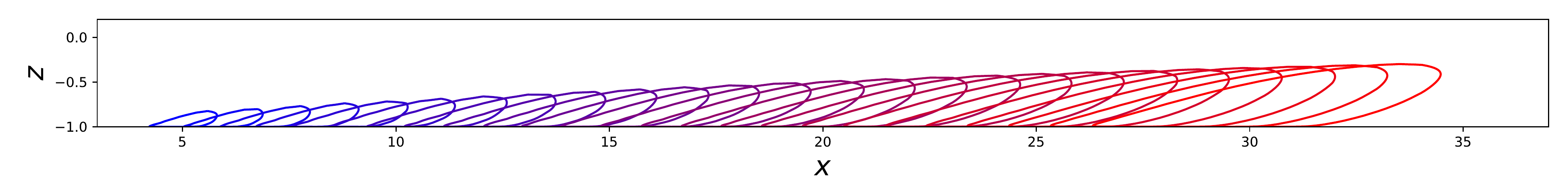}\\
    \includegraphics[width=0.98\textwidth]{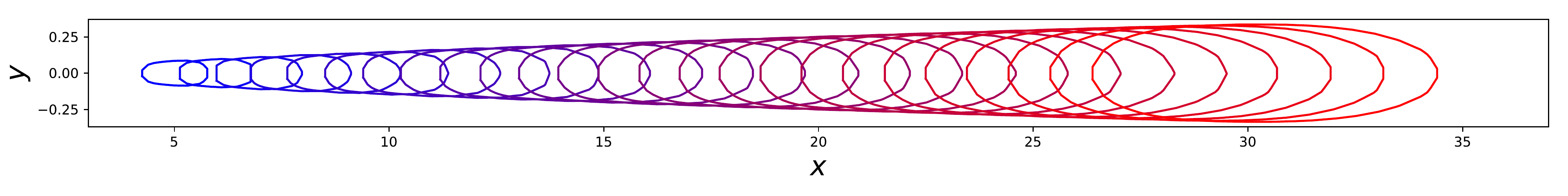}\\
    \includegraphics[width=0.49\textwidth]{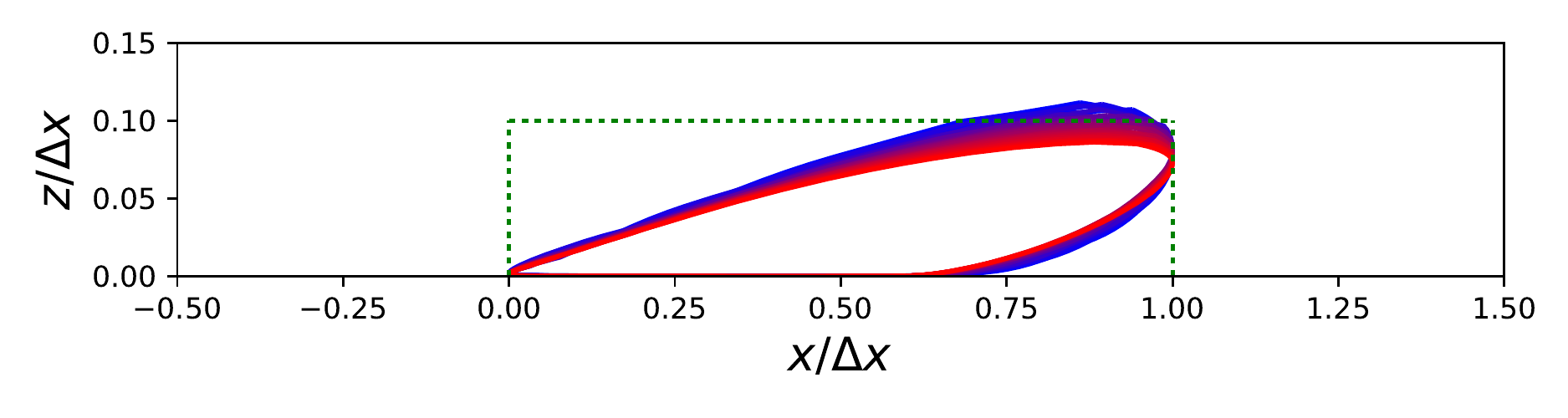}
    \includegraphics[width=0.49\textwidth]{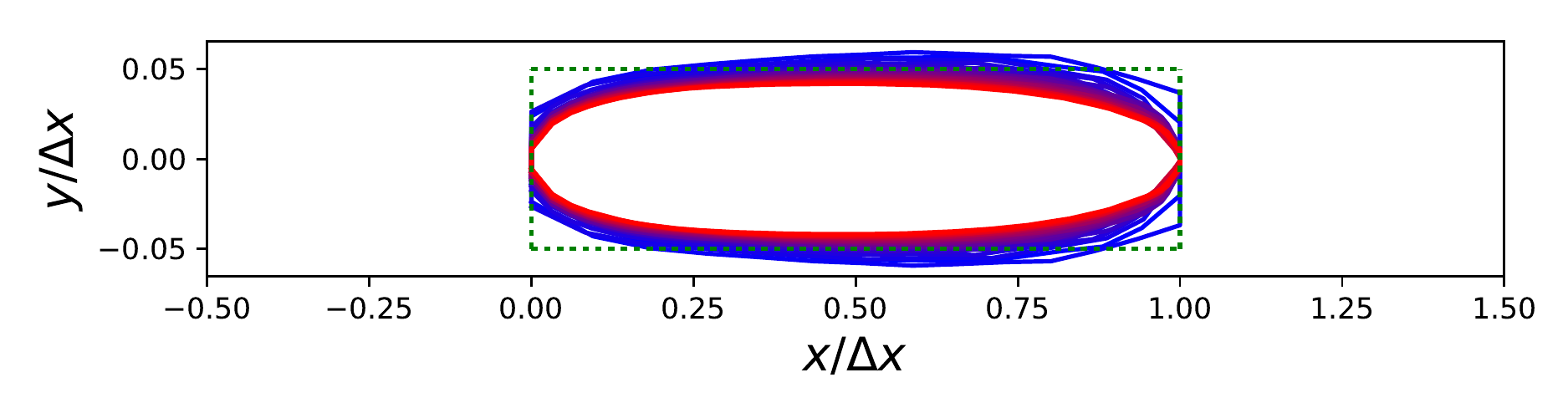}
    \caption{Isocontours of streamwise velocity ($25\%$ of maximum) projected on the $y=0$ plane (top) and on the $z=0$ plane (middle) at times $tU_{CL}/h$ from $8.52$ to $42.7$ in steps of $1.42$; figures on the bottom show these contours scaled by streamwise size ($\Delta x$) and shifted to $x=0$, projected on $y=0$ (bottom left) and $z=0$ (bottom right). Contours are colored to show evolution in time in the overlaid contours in the bottom plots. Rectangles of aspect ratio $10$ are also plotted (dotted lines) to highlight the aspect ratio of the structures.\label{fig:isocontours-I30x}}
\end{figure}

The self-similarity of the coherent structures can be seen more clearly by projecting isosurfaces of streamwise velocity (at $25\%$ of instantaneous maximum) to the $x-z$ and $x-y$ planes, illustrated in \cref{fig:isocontours-I30x} for the I30x case, which has a single dominant streak. These projections show an approximate collapse in rectangular boxes of aspect ratio $10$ (also shown in \cref{fig:isocontours-I30x}).  

\renewcommand{\tmpFigHeight}{0.17\textheight}
\renewcommand{\tmpFigWidth}{0.365\textwidth}
\newcommand{\insShiftY}{0.05\textheight}
\newcommand{\insShiftX}{0.17\textwidth}
\newcommand{\insHeight}{0.12\textheight}
\newcommand{\insWidth}{0.195\textwidth}
\begin{figure}
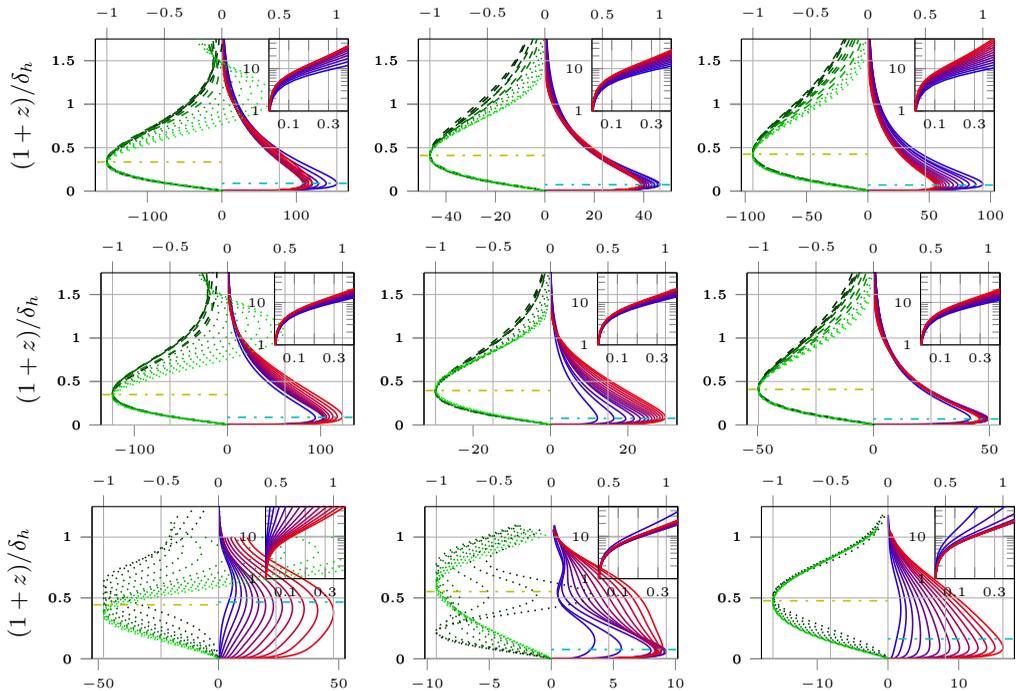

    \centering
    \input{figures/eddyIntRe10000y0p997_Fs_1_0_0} 
    \input{figures/eddyIntRe10000y0p997_Fs_0_0_1}
    \input{figures/eddyIntRe10000y0p997_Fs_0_1_0}\\
    \input{figures/eddyIntRe10000y0p950_Fs_1_0_0}
    \input{figures/eddyIntRe10000y0p950_Fs_0_0_1}
    \input{figures/eddyIntRe10000y0p950_Fs_0_1_0}\\
    \input{figures/eddyIntRe10000y0p500_Fs_1_0_0} 
    \input{figures/eddyIntRe10000y0p500_Fs_0_0_1}
    \input{figures/eddyIntRe10000y0p500_Fs_0_1_0}\\
    \caption{Reynolds stresses for impulses at $z^+=30$ (top row), $z^+=500$ (middle row), and $z=-0.5$ (bottom row) with forcing along $x$ (left columns), $y$ (middle columns), and $z$ (right columns) at times $tU_{CL}/h$ from $5.7$ to $42.7$ in steps of $2.84$; $I_{11}$ with scale on the bottom $x$-axis: blue to red for increasing time, {\color{redBlue}\full}; $I_{13}$ with scale on the top $x$-axis: black to green for increasing time, {\color{midGreen}\dashed}~ for structures with $\delta_h \leq 0.5$ and {\color{midGreen}\dotted}~ otherwise. The average locations of the peaks of $I_{11}$ ({\color{cyan}\dashdot}) and $I_{13}$ ({\color{yellow}\dashdot})are also shown. All curves are truncated to ignore the top half of the channel ($z>1$). Insets show near-wall streamwise energy: $I_{11}$/max($I_{11}$) on the $x$-axis against $z^+$ on the $y$-axis in log scale; note that $z^+$ here is not scaled by $\delta_h$.\label{fig:stress-instant-multiple}}
\end{figure}

The self-similarity is further probed by considering the Reynolds stresses scaled by the height of the coherent structures (distance from the wall). For the periodic domain, the Reynolds stresses are defined as
\begin{equation}
    I_{ij}\big(\frac{1+z}{\delta_h},t\big) := \frac{K}{u^2_\tau}\frac{k_{x,0}}{2\pi} \frac{k_{y,0}}{2\pi} 
            \int\limits_{x=0}^{2\pi/k_{x,0}}\int\limits_{y=0}^{2\pi/k_{y,0}} u_i(x,y,\frac{1+z}{\delta_h},t) u_j (x,y,\frac{1+z}{\delta_h},t) \mbox{d}x \mbox{d}y.
\end{equation}
Here, all wall-normal distances are referenced from the wall (hence the $(1+z)$ dependence). The distances are scaled by $\delta_h$, which represents the wall-normal height of the coherent structure, and is defined as the maximum height from the bottom wall of the $|u|=0.25|u|_{max}$ isosurface. The height of the structure is set to 1 whenever the isosurface extends into the top half of the channel. The constant $K$ is chosen such that the maximum of $-I_{13}$ is 1. This scaling of the Reynolds stresses is similar to the eddy intensity functions of \cite{townsend1976structure}, which is reasonable considering that the coherent structures in the impulse response are reminiscent of the attached eddies in Townsend's attached eddy hypothesis.

Reynolds stresses are shown for the nine cases of impulsive forcing in \cref{fig:stress-instant-multiple}. For all cases except O05x, the coherent structures have non-zero streamwise energy at $z^+ \sim 10$ (shown in inset), with $I_{11}/$max($I_{11}$) being greater than $10\%$, thus showing that they are all attached to the wall. The shear intensity, $I_{13}$, is consistently negative, except for very tall structures with $\delta_h>0.5$ in the core region of the channel. The locations of the peaks of $I_{13}$ show good collapse, and this peak is located further away from the wall than the peak of $I_{11}$. The peaks of $I_{11}$ also show good collapse in the peak locations, but not in the peak magnitudes. The similarities in these three aspects --- wall-attachedness, negative $I_{13}$, and the relative location of peaks of $I_{13}$ and $I_{11}$ --- amongst the different impulse cases demonstrates that there is an underlying structure to the topology of the vortex-streak structures. 

\begin{figure}
    \centering
        \input{figures/energyDecayTime_Re1E4_I30}
        \input{figures/energyDecayTime_Re1E4_I500}
        \input{figures/energyDecayTime_Re1E4_O05}
    \caption{Evolution of total ({\color{red} \dashed}) and streamwise ({\color{blue} \dotted}) kinetic energy density for impulses introduced at different locations: $z^+ \approx 30$ (left), $z^+ \approx 500$ (middle), and $z = -0.5$ (right). For each location, the plots show energy due to impulses along streamwise ($+$), spanwise ($\times$), and wall-normal ($\triangledown$) directions. Markers correspond to times where coherent structures are plotted in \cref{fig:swirl-lapse-full}.\label{fig:energy-multiple}}
\end{figure}
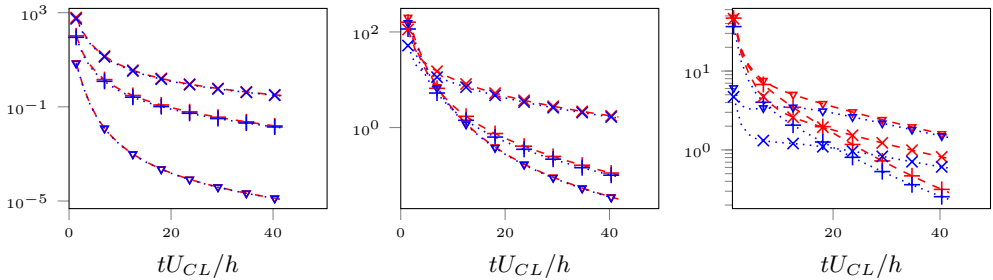

While the coherent structures exhibit self-similarity, they also decay quickly as shown in \cref{fig:energy-multiple}, where the streamwise and total kinetic energy density are shown for all nine impulses. The kinetic energy density is defined for each velocity component, $E_{u_i}$, as
\begin{equation}
    E_{u_i} := \frac{k_{x,0}}{2\pi}\cdot\frac{1}{2}\cdot\frac{k_{y,0}}{2\pi} 
            \int\limits_{x=0}^{2\pi/k_{x,0}}\int\limits_{y=0}^{2\pi/k_{y,0}} \int\limits_{z=-1}^{1} 
            \bigg(\frac{u_i}{u_\tau}\bigg)^2 \mbox{d}x \mbox{d}y \mbox{d}z   ,
\end{equation}
with the total kinetic energy density $E$ being their sum, $E = E_u + E_v + E_w$. For the I30 and I500 cases, the energy in the perturbations reduces by a few orders of magnitude by the time the structures reach the core of the channel ($\delta_h \gtrapprox 0.5$). For the near-wall case, the total energy is predominantly due to streamwise motion for the range of times considered here. For the impulses further away from the wall, however, there is significant energy in the cross-stream components. As the location of the impulse moves further away from the wall, the energy in the cross-stream components remains significant quite late into the decay. We reason that these differences arise due to the local differences in the mean shear and the eddy viscosity; due to the large mean shear near the wall, cross-stream velocities extract a large streamwise perturbation velocity from the mean velocity, while the cross-stream components themselves dissipate due to the eddy viscosity. For impulses away from the wall, the mean shear is significantly weaker, leading to slower extraction of streamwise velocity perturbation energy from the mean velocity. A detailed investigation of the early time evolution of these perturbations is needed to shed light on these mechanisms. 

\renewcommand{\tmpFigHeight}{0.2\textheight}
\renewcommand{\tmpFigWidth}{0.37\textwidth}

%% file: figures/energyDecayTime_Re1E4_I30.tex
\begin{tikzpicture}

\begin{axis}[
xmin=0, xmax=50.6,
ymin=4.74852818768411e-06, ymax=1519.87054229453,
xmode=linear,
ymode=log,
xlabel={\footnotesize $tU_{CL}/h$},
mark repeat={4},
width=\tmpFigWidth,
height=\tmpFigHeight,
tick align=outside,
tick pos=left,
x grid style={white!69.019607843137251!black},
y grid style={white!69.019607843137251!black}
]
\addplot [semithick, red, dashed, mark=+, mark size=3, mark options={solid}, forget plot]
table {%
1.39054419035446 103.870504078375
2.78108838070891 17.8889158487386
4.17163257106337 5.84079588900117
5.56217676141782 2.64560620030209
6.95272095177228 1.44029006573388
8.34326514212673 0.881824190369343
9.73380933248119 0.58566802956333
11.1243535228356 0.412817348166642
12.5148977131901 0.304425667191244
13.9054419035446 0.232549639181062
15.295986093899 0.182706328685112
16.6865302842535 0.146852061935901
18.0770744746079 0.120261940594129
19.4676186649624 0.100028557623955
20.8581628553168 0.0842913578351608
22.2487070456713 0.0718190795155428
23.6392512360257 0.0617727574642669
25.0297954263802 0.0535658852774748
26.4203396167346 0.0467790368229248
27.8108838070891 0.0411060597144573
29.2014279974436 0.0363193524215539
30.591972187798 0.0322468604002028
31.9825163781525 0.0287565296517182
33.3730605685069 0.025745594221084
34.7636047588614 0.0231330564238191
36.1541489492158 0.0208543375389523
37.5446931395703 0.0188573997839498
38.9352373299247 0.0170999039081823
40.3257815202792 0.0155470966820195
41.7163257106337 0.0141702215665881
};
\addplot [semithick, blue, dotted, mark=+, mark size=3, mark options={solid}, forget plot]
table {%
1.39054419035446 89.7684906290683
2.78108838070891 15.1881658014939
4.17163257106337 4.9467637623466
5.56217676141782 2.23664520704058
6.95272095177228 1.2162873700136
8.34326514212673 0.744275435349887
9.73380933248119 0.494318294476871
11.1243535228356 0.348616659228258
12.5148977131901 0.257352532884421
13.9054419035446 0.196891449054787
15.295986093899 0.154994460230286
16.6865302842535 0.124870169873792
18.0770744746079 0.102532955568002
19.4676186649624 0.0855327279698435
20.8581628553168 0.0723032481863173
22.2487070456713 0.0618092122112333
23.6392512360257 0.0533460296053775
25.0297954263802 0.0464217525552624
26.4203396167346 0.0406850314480292
27.8108838070891 0.035879728173079
29.2014279974436 0.0318156547660915
30.591972187798 0.028349198955621
31.9825163781525 0.0253702428002562
33.3730605685069 0.0227931580823091
34.7636047588614 0.0205504915958802
36.1541489492158 0.0185884804462148
37.5446931395703 0.0168638040859094
38.9352373299247 0.0153412076592342
40.3257815202792 0.0139917383927695
41.7163257106337 0.0127914208721177
};
\addplot [semithick, red, dashed, mark=triangle, mark size=1.6, mark options={solid,rotate=180}, forget plot]
table {%
1.39054419035446 7.30838402465818
2.78108838070891 0.715712502571734
4.17163257106337 0.122339950947182
5.56217676141782 0.0334367462579536
6.95272095177228 0.0123564469246474
8.34326514212673 0.0055537741636918
9.73380933248119 0.00285838337563333
11.1243535228356 0.00162429918485426
12.5148977131901 0.000995325737882628
13.9054419035446 0.000647121164338113
15.295986093899 0.000441260670678944
16.6865302842535 0.000312873943626033
18.0770744746079 0.0002291787525048
19.4676186649624 0.000172546045557653
20.8581628553168 0.000132988110947615
22.2487070456713 0.00010458963527875
23.6392512360257 8.37107983862477e-05
25.0297954263802 6.80361516733772e-05
26.4203396167346 5.60490962028042e-05
27.8108838070891 4.67302555469359e-05
29.2014279974436 3.93784957447746e-05
30.591972187798 3.35014975419582e-05
31.9825163781525 2.87470841492814e-05
33.3730605685069 2.48590281467647e-05
34.7636047588614 2.16480334448149e-05
36.1541489492158 1.8972273870399e-05
37.5446931395703 1.67241129442503e-05
38.9352373299247 1.48208887604362e-05
40.3257815202792 1.31984214323234e-05
41.7163257106337 1.18063750506358e-05
};
\addplot [semithick, blue, dotted, mark=triangle, mark size=1.6, mark options={solid,rotate=180}, forget plot]
table {%
1.39054419035446 6.92885335072586
2.78108838070891 0.690031799909261
4.17163257106337 0.118275088554945
5.56217676141782 0.0323461798947743
6.95272095177228 0.0119598940174052
8.34326514212673 0.00537858044333005
9.73380933248119 0.00276979665934157
11.1243535228356 0.00157484891959341
12.5148977131901 0.000965560599198095
13.9054419035446 0.000628111885985349
15.295986093899 0.00042852832869629
16.6865302842535 0.000304006350646804
18.0770744746079 0.000222798781537191
19.4676186649624 0.000167828264501642
20.8581628553168 0.000129416932572037
22.2487070456713 0.000101831411937495
23.6392512360257 8.15428942390603e-05
25.0297954263802 6.63059660783083e-05
26.4203396167346 5.46495277405384e-05
27.8108838070891 4.5584539630537e-05
29.2014279974436 3.84305686922669e-05
30.591972187798 3.27097176352949e-05
31.9825163781525 2.80800470190117e-05
33.3730605685069 2.42927206758475e-05
34.7636047588614 2.11638619196631e-05
36.1541489492158 1.85556746325414e-05
37.5446931395703 1.63635582390193e-05
38.9352373299247 1.45071652061068e-05
40.3257815202792 1.29240994539726e-05
41.7163257106337 1.15654139842324e-05
};
\addplot [semithick, red, dashed, mark=x, mark size=3, mark options={solid}, forget plot]
table {%
1.39054419035446 624.028514807664
2.78108838070891 141.747355786814
4.17163257106337 51.5527863895812
5.56217676141782 25.0638494425686
6.95272095177228 14.434634666989
8.34326514212673 9.27026274616713
9.73380933248119 6.42150577229181
11.1243535228356 4.70149070355609
12.5148977131901 3.59030110016766
13.9054419035446 2.83369524888751
15.295986093899 2.29642001301438
16.6865302842535 1.90156545574596
18.0770744746079 1.60296534091534
19.4676186649624 1.37163671703957
20.8581628553168 1.18868406071312
22.2487070456713 1.04138410964971
23.6392512360257 0.920927694850307
25.0297954263802 0.821065356176546
26.4203396167346 0.737267278377208
27.8108838070891 0.666187370856499
29.2014279974436 0.605311241706199
30.591972187798 0.552720466478587
31.9825163781525 0.506931061220156
33.3730605685069 0.466780387235235
34.7636047588614 0.431347037473371
36.1541489492158 0.399893227394649
37.5446931395703 0.371822266156319
38.9352373299247 0.346647318038405
40.3257815202792 0.323967837977471
41.7163257106337 0.303451722689824
};
\addplot [semithick, blue, dotted, mark=x, mark size=3, mark options={solid}, forget plot]
table {%
1.39054419035446 546.602670836893
2.78108838070891 128.207217345255
4.17163257106337 47.0164151179613
5.56217676141782 22.9678616486163
6.95272095177228 13.276425258776
8.34326514212673 8.55311966582209
9.73380933248119 5.94120217037657
11.1243535228356 4.36087252675199
12.5148977131901 3.33803914916852
13.9054419035446 2.64043306545948
15.295986093899 2.14429282537252
16.6865302842535 1.779144466881
18.0770744746079 1.50263224302996
19.4676186649624 1.2881354311199
20.8581628553168 1.11828069433121
22.2487070456713 0.98135857509972
23.6392512360257 0.869254653269703
25.0297954263802 0.776207571938628
26.4203396167346 0.698038038079437
27.8108838070891 0.631656626993994
29.2014279974436 0.574740185763301
30.591972187798 0.525515027217421
31.9825163781525 0.482608272213204
33.3730605685069 0.444943675196184
34.7636047588614 0.411667801656842
36.1541489492158 0.382096926141075
37.5446931395703 0.355677780458255
38.9352373299247 0.331958728496618
40.3257815202792 0.310568001653292
41.7163257106337 0.291197207380598
};
\end{axis}

\end{tikzpicture}

%% file: figures/energyDecayTime_Re1E4_I500.tex
\begin{tikzpicture}

\begin{axis}[
xlabel={\footnotesize $tU_{CL}/h$},
xmin=0, xmax=49.4495236530964,
ymin=0.0201871799354956, ymax=310.639217696394,
xmode=linear,
ymode=log,
mark repeat={4},
width=\tmpFigWidth,
height=\tmpFigHeight,
tick align=outside,
tick pos=left,
x grid style={white!69.019607843137251!black},
y grid style={white!69.019607843137251!black}
]
\addplot [semithick, red, dashed, mark=+, mark size=3, mark options={solid}, forget plot]
table {%
1.39054419035446 160.138104265951
2.78108838070891 50.5248190526567
4.17163257106337 21.1587298212818
5.56217676141782 11.044331600806
6.95272095177228 6.61621522493601
8.34326514212673 4.3480320372516
9.73380933248119 3.05212728152497
11.1243535228356 2.24986636436281
12.5148977131901 1.72175854036045
13.9054419035446 1.35685920637552
15.295986093899 1.0946608398015
16.6865302842535 0.900087719249541
18.0770744746079 0.751762906235863
19.4676186649624 0.63610006573773
20.8581628553168 0.544150882349462
22.2487070456713 0.46983732910639
23.6392512360257 0.408918197897036
25.0297954263802 0.358362449718393
26.4203396167346 0.315957548096332
27.8108838070891 0.280057024488994
29.2014279974436 0.249414401516061
30.591972187798 0.223071104101233
31.9825163781525 0.200279327642113
33.3730605685069 0.180448160894759
34.7636047588614 0.163105114936738
36.1541489492158 0.147868333005338
37.5446931395703 0.134426199405466
38.9352373299247 0.122522174307331
40.3257815202792 0.111943371193897
41.7163257106337 0.102511849940795
};
\addplot [semithick, blue, dotted, mark=+, mark size=3, mark options={solid}, forget plot]
table {%
1.39054419035446 115.232148512752
2.78108838070891 37.24854696712
4.17163257106337 16.231893050408
5.56217676141782 8.67291273832166
6.95272095177228 5.27543369174732
8.34326514212673 3.50521994412105
9.73380933248119 2.48173430004735
11.1243535228356 1.84258257617227
12.5148977131901 1.41903555950765
13.9054419035446 1.12482224431261
15.295986093899 0.91247084961905
16.6865302842535 0.754268818852096
18.0770744746079 0.63323260802276
19.4676186649624 0.538520636727054
20.8581628553168 0.462967601578259
22.2487070456713 0.40169335387984
23.6392512360257 0.351285414708109
25.0297954263802 0.309301014897395
26.4203396167346 0.273954778526623
27.8108838070891 0.243916705361745
29.2014279974436 0.218178994590329
30.591972187798 0.195965950823152
31.9825163781525 0.176671881639664
33.3730605685069 0.159817697247829
34.7636047588614 0.145019878599131
36.1541489492158 0.131968059362923
37.5446931395703 0.120408575484718
38.9352373299247 0.110132232652571
40.3257815202792 0.100965095581932
41.7163257106337 0.0927614699348006
};
\addplot [semithick, red, dashed, mark=triangle, mark size=1.6, mark options={solid,rotate=180}, forget plot]
table {%
1.39054419035446 200.414847028182
2.78108838070891 69.181326004522
4.17163257106337 27.794946955078
5.56217676141782 13.1837208849477
6.95272095177228 7.06731206107999
8.34326514212673 4.15119306144429
9.73380933248119 2.61588019957772
11.1243535228356 1.74241009914645
12.5148977131901 1.21368833261409
13.9054419035446 0.877026734811707
15.295986093899 0.653449799926502
16.6865302842535 0.499613701641617
18.0770744746079 0.390511705144382
19.4676186649624 0.311088939247526
20.8581628553168 0.251942468629995
22.2487070456713 0.207008244086076
23.6392512360257 0.172263869564603
25.0297954263802 0.144973829812651
26.4203396167346 0.123235957833599
27.8108838070891 0.105700877326896
29.2014279974436 0.091394025477939
30.591972187798 0.0795999209249857
31.9825163781525 0.0697854544671942
33.3730605685069 0.0615479685528175
34.7636047588614 0.0545795244047312
36.1541489492158 0.0486419053091593
37.5446931395703 0.043548869027365
38.9352373299247 0.0391533769919348
40.3257815202792 0.035338269451954
41.7163257106337 0.0320094109478593
};
\addplot [semithick, blue, dotted, mark=triangle, mark size=1.6, mark options={solid,rotate=180}, forget plot]
table {%
1.39054419035446 153.755130038937
2.78108838070891 61.1686780798504
4.17163257106337 25.5622578821885
5.56217676141782 12.3288892171298
6.95272095177228 6.67004030470705
8.34326514212673 3.94092144655819
9.73380933248119 2.49368587294342
11.1243535228356 1.66622278648272
12.5148977131901 1.16350710146443
13.9054419035446 0.842490508075159
15.295986093899 0.6288122952644
16.6865302842535 0.481505552579087
18.0770744746079 0.376863750640542
19.4676186649624 0.300579797540139
20.8581628553168 0.243699573037509
22.2487070456713 0.200438453189952
23.6392512360257 0.166953636466455
25.0297954263802 0.140628309181489
26.4203396167346 0.119640722871659
27.8108838070891 0.102697197077189
29.2014279974436 0.0888625118666949
30.591972187798 0.0774494767008642
31.9825163781525 0.0679456567983493
33.3730605685069 0.0599637267158543
34.7636047588614 0.0532072716420604
36.1541489492158 0.0474468438855463
37.5446931395703 0.0425029499543963
38.9352373299247 0.0382337998193772
40.3257815202792 0.0345263552805807
41.7163257106337 0.0312897466213013
};
\addplot [semithick, red, dashed, mark=x, mark size=3, mark options={solid}, forget plot]
table {%
1.39054419035446 112.714944353579
2.78108838070891 42.3040280811685
4.17163257106337 25.1407913193365
5.56217676141782 18.6752209322017
6.95272095177228 15.0519031937445
8.34326514212673 12.5771210145834
9.73380933248119 10.7320971895462
11.1243535228356 9.29572834958285
12.5148977131901 8.14795871370105
13.9054419035446 7.21347164504738
15.295986093899 6.44110950261108
16.6865302842535 5.79450515195729
18.0770744746079 5.24703979023279
19.4676186649624 4.77881533560504
20.8581628553168 4.37470801857044
22.2487070456713 4.02305655793877
23.6392512360257 3.71475664643269
25.0297954263802 3.44261874554852
26.4203396167346 3.20090589177762
27.8108838070891 2.98499584678752
29.2014279974436 2.79113109472169
30.591972187798 2.61623083511127
31.9825163781525 2.45774973248345
33.3730605685069 2.31356905949777
34.7636047588614 2.18191292140237
36.1541489492158 2.06128302216063
37.5446931395703 1.95040745530973
38.9352373299247 1.84820019532867
40.3257815202792 1.75372882035567
41.7163257106337 1.6661886161089
};
\addplot [semithick, blue, dotted, mark=x, mark size=3, mark options={solid}, forget plot]
table {%
1.39054419035446 51.9098208989366
2.78108838070891 20.9602126338618
4.17163257106337 15.0216346566894
5.56217676141782 12.8816132641713
6.95272095177228 11.333722218026
8.34326514212673 10.001770434784
9.73380933248119 8.84872428605612
11.1243535228356 7.86152229108799
12.5148977131901 7.02120891152971
13.9054419035446 6.30616554808184
15.295986093899 5.69580579993111
16.6865302842535 5.1721678753986
18.0770744746079 4.72023395811373
19.4676186649624 4.32770334867318
20.8581628553168 3.98459266710613
22.2487070456713 3.68281723287389
23.6392512360257 3.41582251967976
25.0297954263802 3.17827596518538
26.4203396167346 2.96581882968527
27.8108838070891 2.77486917939615
29.2014279974436 2.60246673225859
30.591972187798 2.44615007938122
31.9825163781525 2.30386113813342
33.3730605685069 2.17386887105785
34.7636047588614 2.05470908732993
36.1541489492158 1.94513651559267
37.5446931395703 1.84408646410845
38.9352373299247 1.75064399652448
40.3257815202792 1.66401902322274
41.7163257106337 1.58352606976203
};

\end{axis}

\end{tikzpicture}

%% file: figures/energyDecayTime_Re1E4_O05.tex
\begin{tikzpicture}

\begin{axis}[
xmin=1.17308297582, xmax=49.4495236530964,
ymin=0.180457339963009, ymax=62.1972305095751,
xmode=linear,
ymode=log,
mark repeat={4},
xlabel={\footnotesize $tU_{CL}/h$},
width=\tmpFigWidth,
height=\tmpFigHeight,
tick align=outside,
tick pos=left,
x grid style={white!69.019607843137251!black},
y grid style={white!69.019607843137251!black}
]
\addplot [semithick, red, dashed, mark=+, mark size=3, mark options={solid}, forget plot]
table {%
1.39054419035446 46.8878943483031
2.78108838070891 18.4907784052169
4.17163257106337 11.4634019056374
5.56217676141782 8.46478502264418
6.95272095177228 6.75992243047217
8.34326514212673 5.60313453993136
9.73380933248119 4.73314008477696
11.1243535228356 4.04095126867525
12.5148977131901 3.47322462674805
13.9054419035446 2.99958039945432
15.295986093899 2.60035883897959
16.6865302842535 2.26159392800984
18.0770744746079 1.97273773769533
19.4676186649624 1.72551249473793
20.8581628553168 1.5132643300732
22.2487070456713 1.33055867706382
23.6392512360257 1.17290847667615
25.0297954263802 1.03657903329989
26.4203396167346 0.918442718364531
27.8108838070891 0.815867556577618
29.2014279974436 0.726629994166039
30.591972187798 0.648845587562655
31.9825163781525 0.580912727713948
33.3730605685069 0.521467580220452
34.7636047588614 0.469346409187193
36.1541489492158 0.423554362685068
37.5446931395703 0.383240146411975
38.9352373299247 0.347673902947777
40.3257815202792 0.316229500649488
41.7163257106337 0.288368790183147
};
\addplot [semithick, blue, dotted, mark=+, mark size=3, mark options={solid}, forget plot]
table {%
1.39054419035446 36.5735404537353
2.78108838070891 13.3954851164122
4.17163257106337 7.62142743844376
5.56217676141782 5.25097898676519
6.95272095177228 4.01882929911077
8.34326514212673 3.26768691368946
9.73380933248119 2.75211946623696
11.1243535228356 2.3667194021387
12.5148977131901 2.06141750013426
13.9054419035446 1.81022244986878
15.295986093899 1.59844194547746
16.6865302842535 1.41706115735908
18.0770744746079 1.26011230017109
19.4676186649624 1.12337179598395
20.8581628553168 1.00367938818675
22.2487070456713 0.898563301686555
23.6392512360257 0.806023878621576
25.0297954263802 0.724402521802157
26.4203396167346 0.652298650883897
27.8108838070891 0.58851461136032
29.2014279974436 0.532017416865289
30.591972187798 0.481910971372437
31.9825163781525 0.437414963362524
33.3730605685069 0.397848350185086
34.7636047588614 0.362615853099783
36.1541489492158 0.331196695312286
37.5446931395703 0.303135128683207
38.9352373299247 0.278032154803584
40.3257815202792 0.255538459545411
41.7163257106337 0.235348132061803
};
\addplot [semithick, red, dashed, mark=triangle, mark size=1.6, mark options={solid,rotate=180}, forget plot]
table {%
1.39054419035446 47.6908258098119
2.78108838070891 19.2813474257097
4.17163257106337 12.0746838119213
5.56217676141782 9.10713331798291
6.95272095177228 7.55750209313646
8.34326514212673 6.60797924598536
9.73380933248119 5.94968886075284
11.1243535228356 5.44785574551486
12.5148977131901 5.03813231148617
13.9054419035446 4.68768062642916
15.295986093899 4.37873292026756
16.6865302842535 4.10109998985912
18.0770744746079 3.84856534859613
19.4676186649624 3.61707741657792
20.8581628553168 3.40381034393421
22.2487070456713 3.20666003244562
23.6392512360257 3.0239702022504
25.0297954263802 2.85437586318115
26.4203396167346 2.69671311576217
27.8108838070891 2.54996527089911
29.2014279974436 2.4132294945475
30.591972187798 2.28569525582667
31.9825163781525 2.16663040854638
33.3730605685069 2.05536798752429
34.7636047588614 1.95130269329085
36.1541489492158 1.85388029619579
37.5446931395703 1.76259485460416
38.9352373299247 1.67698348508557
40.3257815202792 1.59662240102906
41.7163257106337 1.52112338085148
};
\addplot [semithick, blue, dotted, mark=triangle, mark size=1.6, mark options={solid,rotate=180}, forget plot]
table {%
1.39054419035446 6.09007312596091
2.78108838070891 3.66112890214082
4.17163257106337 3.3072095229311
5.56217676141782 3.31442858002995
6.95272095177228 3.39520319944195
8.34326514212673 3.46691389171565
9.73380933248119 3.50378351507592
11.1243535228356 3.50135663240024
12.5148977131901 3.46367809315552
13.9054419035446 3.39761963906025
15.295986093899 3.31030741106551
16.6865302842535 3.20806520065813
18.0770744746079 3.09609458717091
19.4676186649624 2.97848480717943
20.8581628553168 2.85835222359902
22.2487070456713 2.73801416918972
23.6392512360257 2.61915442858106
25.0297954263802 2.5029655000079
26.4203396167346 2.39026482910136
27.8108838070891 2.28158689354579
29.2014279974436 2.17725483304118
30.591972187798 2.07743551646627
31.9825163781525 1.9821816386814
33.3730605685069 1.89146320000151
34.7636047588614 1.80519214667537
36.1541489492158 1.72323985703366
37.5446931395703 1.64545100131238
38.9352373299247 1.57165350918212
40.3257815202792 1.50166601578331
41.7163257106337 1.43530332406617
};
\addplot [semithick, red, dashed, mark=x, mark size=3, mark options={solid}, forget plot]
table {%
1.39054419035446 45.965681820097
2.78108838070891 16.9628212807197
4.17163257106337 9.46359537875817
5.56217676141782 6.34051813916428
6.95272095177228 4.74260576668247
8.34326514212673 3.8259383027049
9.73380933248119 3.2551735401293
11.1243535228356 2.87311098488733
12.5148977131901 2.59889725442786
13.9054419035446 2.38870053687286
15.295986093899 2.2179999174724
16.6865302842535 2.07278947006565
18.0770744746079 1.94493880401041
19.4676186649624 1.82965658538357
20.8581628553168 1.72406897800547
22.2487070456713 1.62641049229235
23.6392512360257 1.53555920685918
25.0297954263802 1.45076681337212
26.4203396167346 1.37150174818538
27.8108838070891 1.29735793850736
29.2014279974436 1.22800196214931
30.591972187798 1.16314277799975
31.9825163781525 1.1025149395231
33.3730605685069 1.0458685425407
34.7636047588614 0.992965639505417
36.1541489492158 0.943577548988941
37.5446931395703 0.897483379596389
38.9352373299247 0.854471377218371
40.3257815202792 0.814337779608884
41.7163257106337 0.776887993978695
};
\addplot [semithick, blue, dotted, mark=x, mark size=3, mark options={solid}, forget plot]
table {%
1.39054419035446 4.71404319236174
2.78108838070891 2.12601839355538
4.17163257106337 1.5766387723574
5.56217676141782 1.39289164870432
6.95272095177228 1.31443103576409
8.34326514212673 1.27375706334336
9.73380933248119 1.24774464734627
11.1243535228356 1.22660120814519
12.5148977131901 1.20568671254827
13.9054419035446 1.1828582645058
15.295986093899 1.15733553263313
16.6865302842535 1.12909015039516
18.0770744746079 1.09847951600369
19.4676186649624 1.06602711781649
20.8581628553168 1.03229496920851
22.2487070456713 0.997815582350121
23.6392512360257 0.963059308908504
25.0297954263802 0.928422774016802
26.4203396167346 0.894228821958097
27.8108838070891 0.860732234986901
29.2014279974436 0.82812789372091
30.591972187798 0.796559531839502
31.9825163781525 0.766128137461356
33.3730605685069 0.736899504037389
34.7636047588614 0.708910908791827
36.1541489492158 0.682176757022734
37.5446931395703 0.656693322859536
38.9352373299247 0.632442797832973
40.3257815202792 0.609396522181622
41.7163257106337 0.587517738023476
};

\end{axis}

\end{tikzpicture}

%% file: outlook.tex
\section{Conclusion and outlook}\label{sec:outlook}
The impulse response of the eddy-viscosity-enhanced linearized Navier-Stokes equations (eLNSE) shows spatially growing vortex-streak structures that have approximate geometric self-similarity, illustrated here at $Re_\tau = 10000$. The coherent structures produced here are quite similar to the vortex-streak structures known to be important to turbulence \citep{jimenez1991minimal,halcrow2009heteroclinic,hall2010streamwise,hwang2015statistical}. These structures have an aspect ratio (streamwise size to wall-normal size) of approximately $10$, which is close to the experimental observations of aspect ratios of attached eddies in the log layer (reported to be $\approx 14$ based on linear coherence by \cite{baars2017self} for boundary layers). The topology of the vortex streak structure does not change when the forcing is moved from $z^+\approx 30$ to $z^+ \approx 500$, but is dependent on the direction of the impulsive body force. This trend is also observed for forcing at $z = -0.5$, albeit with noticeable differences in the shape of these structures as they extend into the top half of the channel. The coherent structures due to all of the nine impulses considered here evolve to have non-zero energy extending down to $z^+ \sim 10$, i.e., they are all attached to the wall. This includes the cases where the forcing is introduced at $z=-0.5$. The coherent structures for all cases also produce a negative Reynolds shear stress $I_{13}$, which peaks further away from the wall than $I_{11}$. These observations are consistent with the supposed prevalence of attached eddies and the significance of ejection and sweeping events in turbulence. In short, irrespective of the location and direction of the impulse, the resulting structures are vortex-streak structures that are wall-attached and self-similar with negative $I_{13}$ and an aspect ratio of approximately 10. Considering this robust behaviour of the impulse response, we can expect that employing a spatio-temporally distributed body forcing would also produce similar structures. This provides more direct evidence, albeit under the approximation of the eLNSE model, for wall-attached hairpin-like vortices flanking velocity streaks in high Reynolds number flows. 

The energy decay seen in the impulse response needs further clarification, since the large-scale structures appear to have very little energy density, which was also noted by \cite{eitel2015hairpin}. This hurdle could possibly disappear when, instead of response to isolated impulses, an appropriate sum of impulse responses is used; such forcing would represent the response to some spatio-temporally distributed forcing instead of a spatio-temporally localized forcing. Indeed, the Orr-Sommerfeld-Squire operator employed here is capable of producing energy growth that is larger than is needed to capture turbulence statistics \citep[see fig. 16 of][]{zare2017colour}. It is worth noting that the spatio-temporally distributed forcing that must be considered to compensate for the energy decay cannot be the actual Reynolds shear stresses, since a part of these stresses have been absorbed into the eddy viscosity. It would be interesting to determine such forcing, which would be an analogue of the work of \cite{zare2017colour} for a flow-field decomposed into self-similar coherent structures instead of Fourier modes. Unfortunately, no numerical schemes are presently available, as far as the authors are aware, that could achieve this; further limitations arise from the increasing difficulty to obtain spatio-temporally resolved measurements with increasing Reynolds number. 

An alternative approach to constructing turbulence fields with the present coherent structures is the attached eddy model of \cite{perry1995wall}, which describes turbulent fluctuations as a superposition of self-similar attached eddies with a prescribed spatial distribution. The geometrically self-similar vortex-streak structures found in this work have the potential to be used as the building blocks in the attached eddy model; this would bypass the need to prescribe the spatio-temporal distribution of the forcing. However, preliminary calculations show (not included in this paper) that the ratio of the peaks of the streamwise kinetic energy to the negative of the shear stress, max($I_{11}$)/max($-I_{13}$), due to the present coherent structures are higher, by a factor of order $10$, than are needed to match the slopes of the Reynolds stresses in a fully turbulent flow using the attached eddy model. This difference possibly suggests that the required attached eddy velocity field requires more than the linear response considered here. This could include modifications to the linear operator or the inclusion of non-linear interactions. However, these remain open issues that require further investigation. Even so, the robustness of the coherent structures found using the computationally cheap eLNSE model shows good potential in identifying a fundamental self-similar basis to describe turbulence dynamics.